# Will artificial intelligence accelerate or delay the race between nuclear energy technology budgeting and net-zero emissions?

Danish [a,] Adnan khan [b]

[a] School of Management and Marketing, Faculty of Business & Law, Taylor's University, 47500 Subang Jaya, Malaysia

[b] School of Peace and Conflict Studies, University of Peshawar, Peshawar Pakistan

Abstract

This study explores the impact of nuclear energy technology budgeting and artificial intelligence on carbon dioxide ($CO_2$) emissions in 20 OECD economies. Unlike previous research that relied on conventional panel techniques, we utilize the Method of Moment Quantile Regression panel data estimation techniques. This approach provides quantile-specific insights while addressing issues of endogeneity and heteroscedasticity, resulting in a more nuanced and robust understanding of complex relationships. A novel aspect of this research work is introducing the moderating effect of artificial intelligence on the relationship between nuclear energy and $CO_2$ emissions. The results found that the direct impact of artificial intelligence on $CO_2$ emissions is significant, while the effect of nuclear energy technology budgeting is not. Additionally, artificial intelligence moderates the relationship between nuclear energy technology budgeting and $CO_2$ emissions, aiding nuclear energy in reducing carbon emissions across OECD countries. Our findings indicate that transitioning to a low-carbon future is achievable by replacing fossil fuel energy sources with increased integration of artificial intelligence to promote nuclear energy technologies. This study demonstrates that energy innovations can serve as effective climate-resilience strategies to mitigate the impacts of climate change.

## 1. Introduction

In the last 3 decades, rapid global climate change necessitates a global transition to environmentally sustainable energy sources, thereby contributing to reducing carbon emissions. Utilizing renewable energy sources has the potential to decrease global carbon emissions. Several long-term policy instruments are crucial for promoting synergy with renewable energy sources to attain sustained long-term growth (Obekpa and Alola, 2023; Yue et al., 2022). There is a growing agreement that cleaner economic productivity may be achieved through energy efficiency and the effective use of energy forms; this propels the global ambition to achieve carbon neutrality by 2050. Over the next five years, the International Energy Agency (2021) forecasts significant growth in nuclear energy usage, notwithstanding ongoing discussions about its designation as a green investment (International Energy Agency, 2021). Addressing climate change is inherently connected to advancing cost-effective, low-carbon energy technologies that can be implemented globally. Implementing a technology involves and interacts with numerous mechanisms, including economies of scale, research and development, and organizational learning, all of which can contribute to reducing technology costs. Reduced costs create new deployment opportunities, resulting in a beneficial feedback loop or multiplier effect (Jessika Trancik et al., 2015). Nuclear energy is a prime example of this process among low-carbon electricity technologies.

Scholars and policy analysts have taken an interest in nuclear energy, considering these developments. There is an opportunity for global energy systems to reduce their environmental impact by using low-carbon energy sources and technology. Despite environmentalists' longstanding opposition, some have suggested reviewing nuclear energy as a clean option to fight global warming (Hassan et al., 2023, 2022). Many clean energy sources can now be used due to technological advances and lower costs, despite transportation costs, growing commodity prices, and supply chain issues. Despite their popularity, renewable energy sources are being promoted to alleviate the climate catastrophe and boost economic growth. To accelerate energy innovation for decarbonization, public investment in research, development, and demonstration (RD&D) (Meckling et al., 2022; Simionescu, 2023). The budget allocation for research and development (R&D) in renewable energy, energy efficiency, and nuclear energy is critical for achieving sustainable development goals 7, 9, and 13. This will accelerate progress in energy innovation, support the shift to cleaner energy sources, and assist in mitigating climate change (Jin et al., 2023a). Sustainable development cannot be achieved without addressing the energy sector, which must simultaneously reduce harmful emissions while increasing access to energy for more individuals. There are risks associated with every technology that might affect people and the planet. Although low-carbon energy sources do not emit carbon dioxide when they are in use, they do produce emissions and trash throughout their lifecycle, from building to production to decommissioning. Therefore, it is critical to compare potential energy technologies to determine how well they mesh with sustainable development objectives

(World Nuclear Association, 2014). The potential impact of energy sector R&D is becoming more recognized considering rising worries about climate change and global warming. Government support for energy sector R&D is critical for the discovery of low-carbon technology that can maintain a steady supply of energy, lessen reliance on fossil fuels, and mitigate the financial burden of growing nonrenewable energy prices (Danish and Ulucak, 2022).

Artificial intelligence (AI) is a pivotal force driving industrial and technical innovation, significantly impacting advancements in energy conservation and reducing carbon emissions. Enhanced comprehension of the impact of AI on energy transition and carbon emissions may facilitate its application in achieving carbon neutrality (Wang et al., 2024; Zhou et al., 2024). Energy production is only one of several industries that brought about a new digital age thanks to technological advancements. Rapid utilization of nuclear energy is essential to meet the decarbonization targets set for 2050. When licensing nuclear technology for commercialization, digital replicas make nuclear reactor commissioning and full-cycle operation easier. Finding ways to use fewer computing resources is necessary to industrialize nuclear energy and reduce greenhouse gas emissions (GHGE). New digital technologies are opening up exciting possibilities for practically bringing theory and practice together. Innovative technology has reached the industrial level rapidly due to the energy transition. Various technologies, including AI, are part of the nuclear sector. A regulatory framework for the age of digitization can only be established by first identifying the relevant variables. New nuclear technology capabilities will be able to flourish in the future thanks to this framework (Renteria del Toro et al., 2024).

Climate change adaptation and mitigation remain critical to achieving sustainable development while reducing climate vulnerability, particularly among climate-exposed and sensitive regions. Yet, achieving a balance between climate-resilience pathways, high economic productivity, high human development, and energy efficiency appears complex, leading to potential trade-offs (Sarkodie et al., 2023). Against the backdrop and driven by the new possibilities for cleaner manufacturing with AI, this study focuses on how AI and nuclear energy technologies influence environmental performance. Strict environmental restrictions and sustainable nuclear energy generation could be crucial in achieving sustainable development, which aims to balance economic growth, environmental protection, and social well-being while ensuring a sustainable future. This study seeks to address several questions based on the discussion. Firstly, does integrating artificial intelligence with nuclear energy technology contribute to reducing carbon emissions? Secondly, can incorporating artificial intelligence into the nuclear energy technology sector assist in managing carbon emissions? In this paper, we expand the nuclear energy and pollution model by introducing nuclear energy technology and artificial intelligence to evaluate the moderating role of AI in the relationship between NETB and $CO_2$ emissions.

This study contributes significantly to the existing literature. (i) This piece of research work examines the impact of AI and nuclear energy technology on pollution levels in selected OECD countries. This research uses a panel data methodology to elucidate the pollution levels in OECD countries that consume nuclear energy from 1994 to 2020. (ii) The study introduces the interaction effect between AI and nuclear energy technology budgeting to examine AI's moderating role in the relationship between nuclear energy technology budgeting and $CO_2$ emissions. The moderating role is essential for optimizing the deployment and utilization of nuclear energy, as AI-driven strategies can enhance the low-carbon advantages of nuclear energy. (iii) This study employs the Method of Moment Quantile Regression (MM-QR) for empirical estimation. MM-QR utilizes the method of moments to develop more robust estimators compared to traditional quantile regression, facilitating enhanced inference in complex, non-linear data structures where ordinary least squares (OLS) methods may be inadequate. MM-QR contributes methodologically by offering quantile-specific insights while addressing endogeneity and heteroscedasticity, thereby enhancing the understanding of relationships within complex data structures. This tool is essential in empirical research, especially in disciplines where it is crucial to capture distributional heterogeneity and address complex data challenges.

The article organizes the remainder of the section: Section 2 provides information about the profile of sample countries. Afterward, use the econometric tools and data in Section 3, then explain and evaluate the empirical findings in Section 3. In the last section, the paper presents its findings.

## 2. The Profile of Sampled OECD Countries

From 1990 to around 2010, coal and other fossil fuels played a significant role in the continuous emissions rise. However, global $CO_2$ emissions have stabilized over the past decade due to improved energy efficiency and the increasing use of renewable energy sources. This trend is especially evident in the United States and the European Union. The COVID-19 epidemic caused a short-lived decrease in emissions in 2020, but when economies resumed in 2021, they quickly recovered, highlighting the ongoing dependence on fossil fuels in numerous regions worldwide (Hannah Ritchie and Max Roser, 2020). In 2022, OECD nations substantially contributed to global $CO_2$ emissions, mainly from energy-related sources such as coal, oil, and natural gas. Global emissions from the energy sector have continued to rise, particularly due to coal-fired electricity generation. Despite certain regions achieving progress in emissions reductions via the implementation of renewable energy, global emissions have nonetheless surged to an unprecedented level, exceeding 36.8 billion metric tons. Countries within the OECD, especially in Europe and North America, experienced diverse trends. Conversely, the United States experienced a modest rise in emissions attributed to increased demand in the building sector during severe weather events, despite the growth of clean energy sources such as wind and solar (IEA, 2022). Figure 1 shows annual per capita $CO_2$ emissions worldwide in 2024.

======= **INSERT FIGURE 1 HERE** =========

The utilization of nuclear energy has shown a consistent upward trend since the early 1960s. With a 300-gigawatt electrical (GWe) capacity, nuclear energy currently contributes 25% to the OECD countries' electricity consumption. It is crucial for maintaining the energy security of the OECD. Nuclear energy constitutes the most significant proportion of clean energy production among OECD countries. The Paris Agreement of 2015, according to the OECD database, seeks to restrict global warming to below 2°C relative to pre-industrial levels; projections indicate a rise of 3–6 degrees Celsius in the global average temperature by the century's end. Nuclear power plants, expected to supply low-carbon electricity to many countries for the forthcoming decades, will experience significant disruptions in their operations (Nuclear Energy Agency, 2022). As significant economies and notable greenhouse gas emitters, the OECD countries bear a considerable responsibility by investing significantly in energy innovation via research and development (R&D). This study seeks to investigate and propose effective policy measures concerning the role of R&D in the nuclear energy sector to mitigate pollution.

Applications aimed at boosting workplace productivity, particularly in the finance and manufacturing sectors, drove notable growth in the adoption of AI in OECD countries in 2022. A significant number of companies have implemented AI technologies for various functions, including production tracking, image recognition, and natural language processing; however, the extent of adoption has shown considerable variation. Large firms typically spearhead AI integration, particularly for tasks aimed at optimizing operations and enhancing output quality. Nonetheless, the general adoption rate stayed comparatively low, influenced in part by the expenses linked to AI implementation and the insufficient availability of skilled personnel required for its effective operation (Stijn, 2023). The trend for the number of artificial intelligence-related patents in OECD countries comparing the data of 1995 and 2020 is shown in Figure 2.

======= INSERT FIGURE 2 HERE =======

======= INSERT TABLE 1 HERE ========

### 3. Theoretical background and literature review

*3.1. Theoretical background*

Following Fatouros and Stengos, (2023), this study within the framework of an endogenous growth model with environmental constraints and vertical innovations derives optimal nuclear energy technology. Innovation in this study is directed to nuclear technologies and artificial intelligence.

Several fundamental theoretical frameworks inform policy, technological advancement, and strategic planning for achieving net-zero carbon emissions. These frameworks primarily focus on integrating nuclear energy technology and artificial intelligence (AI). The theory of technological innovation underpins this study, emphasizing the role of technological advancements in facilitating societal progress and addressing significant issues such as carbon emissions. Key concepts like the diffusion of innovation illustrate the impact of technologies like nuclear energy and AI on various industries and society. Innovations in nuclear energy technology—such as small modular reactors (SMRs) and advanced reactor designs—present opportunities for improved safety and cost efficiency in the context of achieving net-zero emissions. Meanwhile, AI enhances the optimization of reactor operations, predictive maintenance, and safety monitoring. The combination of advancements in nuclear technology and AI-driven management significantly improves the efficiency and scalability of nuclear energy as a low-emission power source.

Energy transition models analyze historical trends, transformation obstacles, and elements promoting the transition to sustainable energy. Nuclear energy provides consistent baseline power, which can support the intermittent nature of renewable sources. Additionally, AI enhances energy management and grid reliability, facilitating a smoother transition. Policies must promote their adoption, ensure safe operation, and facilitate integration with other low-carbon technologies by enabling nuclear energy and AI to substantially aid in achieving net-zero objectives. Environmental economics shapes the design of incentives and penalties, ensuring the economic viability and competitiveness of nuclear and AI-driven systems within the energy market.

### 3.2. Literature Review

This outline presents a well-organized framework for examining the effects of nuclear energy and AI on emissions reduction, emphasizing diverse viewpoints found in the existing literature. Nuclear energy is a reliable low-carbon power source for mitigating global $CO_2$ emissions. Previous studies examined the relationship between sufficient investment in nuclear energy infrastructure, research, and development and decreased $CO_2$ emissions. The literature on nuclear energy technology and $CO_2$ emissions emphasizes nuclear power's role as a low-carbon energy source that supports the transition to sustainable energy systems. Studies underscore the importance of energy technology innovation and effective policy in achieving emissions reductions and advancing nuclear energy. Jordaan et al., (2017a) highlighted Canada's uneven investment between fossil fuels and clean energy technologies, urging better alignment of federal and regional policies and enhanced financial access for clean energy projects to drive innovation and emissions reductions. This reflects a broader trend across the OECD and other developed nations, where policy plays a critical role in promoting or hindering clean energy investment (Alvarez-herranz et al., 2017; Danish and Ulucak, 2021).

Various studies showed that advancements in energy technologies, such as nuclear, play a crucial role in decreasing pollution. For instance, Álvarez-herránz et al., (2017) discovered that prompt innovation in the energy sector improves environmental quality. Shahbaz et al., (2018) supported this finding in France, while, and Danish and Ulucak, (2021) emphasized the importance of energy R&D in China. Bashir et al., (2024) arrived at comparable conclusions for advanced industrial economies, reinforcing the argument that energy innovation, when supported by sufficient funding and policy measures, has the potential to reduce emissions. Recent studies concentrate on the budgets allocated for nuclear energy research and development (R&D) and their impact on emissions. Uche et al., (2023) observed that government-funded R&D for nuclear and other clean energy technologies offered crucial insights for effective policymaking. (Huang et al., 2024a) found that enhanced nuclear energy budgets positively contribute to environmental quality; however, they recommend customizing policies to meet the specific requirements of each economy. Jin et al., (2023) noted that, although nuclear R&D in Germany has a restricted influence on emissions, it does mitigate ecological degradation, highlighting the complex effects of nuclear energy policies in different nations. Huang et al., (2024b) employed a quantile-based methodology to evaluate the asymmetric effects of nuclear energy budgets on $CO_2$ emissions across major economies. The findings indicated that focused investments in nuclear energy led to a reduction in emissions across particular budget quantiles, highlighting the importance of optimal budget allocations for enhancing the climate advantages of nuclear energy. Shen et al., (2024) noted favorable connections between nuclear technology and emission reduction in Europe while emphasizing the varied national patterns in these relationships. In a related area, Yıldırım et al., (2022) explored the nonlinear effects of environmental innovation on emissions in OECD countries, revealing that innovation leads to a reduction in emissions up to a specific threshold, beyond which "rebound effects" manifest. This finding highlights the critical need for stringent environmental policies to sustain the beneficial effects of innovation, a principle that also extends to nuclear energy. In the absence of sufficient policy backing, the potential environmental advantages of nuclear and other low-carbon technologies might remain unfulfilled (Wang and Zhu, 2020). In summary, the existing literature consistently highlights the potential of nuclear energy to reduce $CO_2$ emissions, particularly when bolstered by focused policy, sufficient investment in research and development, and harmonious integration with renewable energy sources.

## 4. Material and Methods:

### 4.1. Empirical Model Specification

The study's objective is to empirically investigate the nuclear energy, AI, and $CO_2$ emissions nexus. For cross-sectional or panel data analysis (e.g., across OECD countries over multiple years), the model can be specified as follows:

$$\text{Ln}(CO_2)_{it} = \alpha_{it} + \text{Ln}\,\beta_1(NRDD)_{it} + \text{Ln}\,\beta_2(AI)_{it} + \mu_{it} \qquad (1)$$

In the above equation (1). $\beta_1$ and $\beta_2$ capture the effects of nuclear energy and AI on emissions. Nuclear Energy $\beta_1$ negative correlation with $CO_2$ emissions, as nuclear power generally reduces reliance on fossil fuels. *AI* Efficiency Impact $\beta_2$ Expected to be negative, as AI improvements in energy management reduce waste and optimize renewable and nuclear use, lowering emissions. *i* and *t* represent the country and period, respectively

Nuclear energy plants can be made more efficient, more secure, and operationally effective with the help of artificial intelligence (AI), which can moderate the relationship between nuclear energy and $CO_2$ emissions. Since AI-driven techniques can help maximize the low-carbon benefits of nuclear energy, this moderating role is especially critical in maximizing the deployment and utilization of nuclear energy as its facilities can become even more effective in reducing emissions if AI can enhance their performance and efficiency. The optimal operation of nuclear power plants is made possible by artificial intelligence (AI) in predictive maintenance, defect detection, and real-time performance monitoring. This results in reduced downtime and longer plant life. Nuclear power can reduce carbon dioxide emissions by replacing fossil fuels more reliably due to its constant and efficient functioning (IAEA, 2023).

To improve the energy mix's carbon efficiency and decrease the demand for fossil fuel backup energy, AI algorithms can dynamically modulate nuclear production in reaction to changes in renewable generation (IEA, 2022). Emissions reductions are proportional to the degree to which AI enhances the integration of renewables and nuclear energy.

A regression model's interaction term allows one to examine AI's moderating function. For example, we can construct an empirical model in the following manner:

$$\text{Ln}(CO_2)_{it} = \alpha_{it} + \beta_1 \text{Ln}(NRDD)_{it} + \beta_2 \text{Ln}(AI)_{it} + \beta_3 \left[\text{Ln}(AI)_{it} + \text{Ln}(NRDD)_{it}\right] + \mu_{it} \quad (2)$$

In this model, the interaction term ($LnAI_{it} \times LnNRDD_{it}$) would help capture the moderating effect of AI on the relationship between nuclear energy and $CO_2$ emissions. A negative and statistically significant coefficient for this term would suggest that AI enhances the emissions reduction impact of nuclear energy.

### 4.2. *Method of Moment Quantile Regression (MM-QR) method*

For empirical estimation, this study used the Method of Moment Quantile MM-QR method proposed by Sim and Zhou, (2015). Statistics traditionally use the MM-QR estimation technique to find estimates for model parameters by equating sample moments, such as sample means or variances, to their theoretical counterparts. In quantile regression, the MM-QR estimates the quantiles of the response variable based on the predictors. This gives a more complete picture of how the variables are connected than mean regression. The MoM approach allows one to obtain quantile estimates without needing to make strict assumptions about the underlying distribution of residuals, making it suitable for heteroscedastic or non-normally distributed data.

This method provides a comprehensive perspective on the influence of predictors on the distribution of a response variable, extending beyond the mean. This is important in situations where the mean does not sufficiently reflect the data's distribution or when outliers have a substantial impact on mean-based estimates. This method enhances robustness against outliers and model misspecification, rendering it especially beneficial in domains such as finance and economics, where these challenges commonly occur. This method exhibits reduced sensitivity to outliers compared to traditional mean-based regression, as it emphasizes quantiles rather than the entire distribution. The MM-QR method addresses non-normal distributions, effectively accommodating diverse error structures in the data, such as skewed or heavy-tailed distributions. MM-QR offers a comprehensive view of the conditional distribution of the response variable, which is essential for analyzing variability and skewness in data. MM-QR estimates parameters by addressing moment conditions, generally employing optimization methods. In this process, the error between the sample moments and theoretical moments is minimized. After estimating, interpret the regression coefficients as effects on the specified quantile of the response variable. These interpretations can provide insights that differ across quantiles and reveal trends not visible in mean regression. For quantile τ, the MMQR equation can be represented as follows:

$$Q_\tau(CO_2 emission/X = X'\beta(\tau) + \epsilon(\tau) \tag{3}$$

Where $Q_\tau(.)$: The conditional quantile function of $CO_2$ emissions at quantile $\tau$. X is a Vector of independent variables, which includes nuclear energy budgeting, AI investment, and interaction variable (AI*NRDD). $\beta\tau$ Vector of coefficients specific to quantile $\tau$ that reflect the impact of each independent variable on $CO_2$ emissions, and $\epsilon(\tau)$ is the error term associated with quantile $\tau$.

The equation can be expanded to explicitly include NRDD and AI, as follows:

$$Q_\tau(CO_2 emission/NRDD, AI, (AI * NRDD)) = \beta_0(\tau) + X'\beta_1(\tau).NRDD + X'\beta_2(\tau).AI + X'\beta_3(\tau) AI * NRDD + \epsilon(\tau) \quad (4)$$

Where, NRDD Nuclear energy budgeting (investment or budget allocation for nuclear technology), AI number of registered AI-related patents, potentially supporting optimization and efficiency in nuclear energy. AI*NRDD is the interaction between Artificial intelligence and nuclear energy budgeting that may influence $CO_2$ emissions. The MM-QR models are especially valuable in environmental and energy economics for capturing heterogeneous impacts across different levels of emissions, allowing for a more nuanced policy and investment approach. Finally, perform diagnostic checks to validate the model. This may include testing the robustness of the moment conditions, checking for model misspecification, and comparing the quantile regression estimates with other estimation methods.

*4.3. Data*

This paper covers the annual dataset for selected 20 OECD countries (refer to Table 1) from 1994 to 2020. The dependent variable is per capita $CO_2$ emissions (million tons of $CO_2$). Data about $CO_2$ emission is derived from the British Petroleum Statistical Review (BP, 2021). AI and Nuclear energy technology research and development budgets (NRDD) are used as independent variables to explain $CO_2$ emissions. The NRDD is measured for nuclear energy technology budgeting and measured (in a million USD and PPP). The NRDD variable can provide important information for SDG 7. AI is measured as the number of registered AI-related patents. Data on NRDD arrives from the International Energy Agency. AI data is taken from OECD databases. The trend in artificial intelligence, nuclear energy RD&D, and $CO_2$ emissions in sample OECD countries is shown in Figure 3.

======= **INSERT FIGURE 3** =======

## 5. Results

The first step in panel data analysis is to examine cross-sectional dependence (CD). Refraining from ignoring cross-sectional reliance can lead to biased estimates and incorrect conclusions. Therefore, it is advisable to use techniques such as fixed effects with cross-sectional averages, factor models, or spatial econometric models to address this dependency issue effectively. This study employs the CD test for this purpose by Pesaran, (2015). The results in Table 2 reject the null hypothesis, indicating that CD exists. Due to this dependence, shared shocks can influence parameter estimates, potentially resulting in biased outcomes in models that assume unit independence. In international economics, cross-sectional dependence often suggests that global factors—such as commodity prices, financial crises, or technological spillovers—impact multiple countries simultaneously. This understanding informs the selection of appropriate models and policy recommendations.

======= **INSERT TABLE 2 HERE** ========

Once it is proven that the dataset is cross-sectionally dependent, the next step is to use panel unit root tests. These give essential information about the data's structure and help analysts choose the best modeling methods for accurate results. For this purpose, we utilize the CIPS and CADF unit root tests by Pesaran, (2007). The findings in Table 3 indicate a noteworthy test statistic (typically at a 5% significance level), enabling the rejection of the null hypothesis, which suggests that the series is stationary at its first difference. A non-significant statistic indicates that the series is non-stationary.

======= INSERT TABLE 3 HERE ========

Adopting artificial intelligence (AI) positively correlates with $CO_2$ emissions across all quantiles (see Table 4). Generally, an increase in AI adoption leads to an increase in $CO_2$ emissions, irrespective of the examined emission level (low, medium, or high quantiles). This finding suggests that, although AI can improve efficiency and reduce emissions in certain contexts, its deployment might also lead to environmental costs that could outweigh these benefits. A positive coefficient at lower quantiles suggests that, even in nations with relatively low emissions, including Denmark, Switzerland, Norway, Finland, Sweden, Portugal, Hungary, and Austria, an increase in AI adoption is associated with higher $CO_2$ emissions. This could arise from the essential infrastructure and energy needs required for implementing AI technologies in settings that typically show reduced emissions. For average emitters (Belgium, Netherlands, Spain, France, Italy, UK, Australia, and Türkiye), a positive coefficient at the median level (e.g., 50th percentile) suggests a link between AI adoption and increased emissions. This indicates that the efficiencies gained through AI may not adequately offset the emissions generated by data processing, storage, and computational requirements. At upper quantiles (e.g., 75th or 90th percentile), a positive coefficient indicates that even significant emitters (Canada, Germany, Japan, and the USA) see an increase in $CO_2$ emissions with the rise of AI adoption. Industries with significant emissions, including energy, manufacturing, and logistics, could experience a heightened environmental impact from AI due to the substantial computing power and energy consumption required for extensive AI operations.

An insignificant correlation between funding for nuclear energy technology and $CO_2$ emissions across quantiles suggests that variations in funding for nuclear energy technology do not significantly impact CO emissions at different levels of the emissions distribution (see Table 4). This finding may arise from multiple factors about nuclear energy investments and the broader energy production and emissions context. An insignificant relationship across quantiles in quantile regression means this lack of effect is the same for all emissions levels, from lower to higher quantiles. This consistency indicates that irrespective of a country's status as a low or high $CO_2$ emitter, increases in nuclear energy funding are not presently associated with emissions reductions, reinforcing the notion that nuclear budgeting may exert limited or delayed direct effects on $CO_2$ emissions across various emission levels.

There is a negative coefficient for AI's moderating role in the relationship between funding for nuclear energy technology and $CO_2$ emissions across all quantiles. This suggests that AI may help lower the $CO_2$ emissions from nuclear energy projects (see Table 4). AI is a mitigating factor, diminishing the positive relationship between nuclear energy investment and emissions across all levels. The application of AI in nuclear energy development and operations indicates that its efficiencies and optimizations could lead to decreased emissions throughout the project lifecycle. At lower emission levels (10th or 25th percentile), AI may have a negative moderating effect. This means that in places with moderate emissions, AI may reduce the initial CO2 footprint of nuclear energy projects. Figuring out this may necessitate the optimization of construction and logistical procedures, reducing waste, and improvement of resource efficiency. The median quantile, or 50th percentile, for average emitters, shows that AI has a moderating effect that greatly reduces emissions related to nuclear energy projects. This is achieved through AI-driven efficiency, such as predictive maintenance, better energy management, and automation routine tasks. There is a negative coefficient for the moderating effect of AI at high quantiles, like the 75th or 90th percentile. This means that adding AI may have a big effect on situations with high emission levels. Artificial intelligence has the potential to reduce the carbon footprint associated with extensive nuclear initiatives, enhance operational efficiency, and optimize the integration of nuclear facilities into the energy grid.

In quantile regression analysis, plotting the regression results across different quantiles provides a robust check of how the relationship between the independent and dependent variables changes as the outcome variable's distribution changes. Through the graphic representation of these results, we can understand the stability and variation of coefficients, which helps confirm the robustness of the findings. In summary, quantile regression plots serve as an important tool for assessing robustness by demonstrating the variability (or consistency) of coefficient estimates across different quantiles. Consistent coefficients indicate reliability, whereas fluctuations provide understanding of distributional diversity. This visual representation bolsters confidence in the results, especially when dealing with intricate data that may include outliers, non-normal distributions, or endogeneity concerns. By validating the relationship across quantiles, quantile regression plots act as a significant instrument for evaluating the robustness and depth of empirical findings.

Quantile regression plots (see Figure 4) exhibit robustness to outliers and non-normal data distributions. If the plot indicates that the coefficients keep their significance and direction across quantiles, this resilience to outliers further corroborates the stability and validity of the relationship. It provides confidence that the results are not unduly influenced by extreme values or deviations from normality.

======= **INSERT TABLE 4 HERE** ========
======= **INSERT FIGURE 4 HERE** =======
======= **INSERT FIGURE 5 HERE** =======

This approach has been used to estimate Eqs 3 & 4 to check the robustness of the quantitative regression analysis results obtained by the MM-QR method. (1) and (2). Moreover, we conduct a robustness check by applying the alternative method of the augmented Anderson–Hsiao estimator AAH estimator by (Chudik and Pesaran, 2022) and the Panel -Corrected Standard Errors (PCSE) method by (Beck et al., 1995) for comparison purposes. To this end, we include $\ln(CO_2)_{it-1}$ in Eqs. (1) and (2) to estimate the parameters. Here, we should indicate that as the cross-sectional dependence exists in our dataset, using conventional panel data estimators like fixed effect, random effect, or pooled OLS to estimate the static panel data models can lead to biased and inconsistent results. For this reason, we used the AAH regression method which not only considers the cross-sectional dependence but also solves autocorrelation and reduces the endogeneity issue. The robust analysis findings are reported in Table 5 and confirm our results about the sign of the coefficient of NRDD, AI, and the interaction effect of AI and NRDD.

======= **INSERT TABLE 5 HERE** ========

## 4. Discussion

AI directly impacts $CO_2$ emissions positively. First, the possible reasons are that training and deploying AI models, especially large ones, require significant computational resources and, therefore, energy. Data centers, cloud computing, and the hardware powering AI algorithms contribute to $CO_2$ emissions, often more than the efficiencies gained. Second, the infrastructure supporting AI—data centers, sensors, high-performance computers, etc.—demands substantial energy. This infrastructure may lead to a direct increase in emissions, especially if powered by fossil-fuel-based energy sources. Third, AI-enabled innovations can lead to higher productivity, which might drive up consumption and production. This "rebound effect" can result in higher energy demand and emissions, offsetting the efficiencies brought by AI. This positive association across all quantiles signals a potential trade-off: while AI offers benefits, its environmental costs need careful management to avoid inadvertently worsening $CO_2$ emissions on a large scale. Overall, an insignificant relationship across quantiles highlights the complex and potentially indirect role of nuclear energy budgeting in emissions reduction. This finding suggests that while nuclear energy funding is crucial for long-term energy strategy, immediate and observable impacts on $CO_2$ emissions may require additional complementary investments and policy measures to amplify the potential emissions reduction benefits of nuclear energy.

In many countries, nuclear energy is part of a diverse energy mix, and the impact of nuclear budgeting on $CO_2$ emissions may be diluted by the presence of other energy sources, especially if fossil fuels still constitute a large share of the energy supply. As a result, increases in nuclear funding may not have a strong, immediate, or quantifiable effect on emissions, especially if nuclear power generation remains a smaller portion of the overall energy grid. Nuclear energy indirectly reduces $CO_2$ emissions by offsetting reliance on carbon-intensive fossil fuels. However, if nuclear investments are primarily allocated toward safety improvements, decommissioning, or research and development rather than increasing capacity, the immediate effect on $CO_2$ emissions could be minimal or difficult to detect. The quantile regression showing insignificant results across quantiles could suggest that any potential impact of nuclear budgeting on emissions may only become significant above a certain threshold of funding or nuclear capacity, which the current budget levels have not yet reached. This threshold effect could mean that without reaching substantial nuclear capacity expansion, the impact on emissions remains marginal. The creative design of advanced nuclear energy systems, thorough research and development, and widespread use of next-generation advanced nuclear energy technology will increase the economy's competitiveness and enhance the safety of the nuclear energy system (Zhan et al., 2021).

AI can aid in the planning, scheduling, and managing of nuclear plant construction, optimize resource utilization, and mitigate delays. AI can enhance supply chain efficiency and minimize transportation requirements, reducing the indirect $CO_2$ emissions of construction activities. After nuclear plants become operational, AI can improve energy efficiency by precisely aligning energy production with demand, thereby minimizing excess energy production and related emissions. AI algorithms can forecast and mitigate equipment problems before failures, decreasing the necessity for energy-intensive repairs and minimizing potential downtime. This may also prolong the lifespan of equipment, thereby reducing the frequency of replacements and the related manufacturing emissions. Artificial intelligence can facilitate the integration of nuclear energy into the energy grid by balancing supply and demand, optimizing energy storage, and minimizing waste. This mitigates emissions from backup fossil-fuel power sources that may be required for grid stabilization. Artificial intelligence can enhance waste management processes and resource utilization, thereby minimizing material waste and emissions associated with waste disposal or recycling. In sum, this negative moderating effect highlights AI's potential to make nuclear energy development more sustainable. By reducing emissions associated with nuclear projects across all quantiles, AI could play a crucial role in achieving cleaner energy solutions and enhancing the environmental sustainability of nuclear power across various levels of existing emissions.

5. **Conclusion**

The issue of global warming is likely to persist, even beyond the established deadline for reaching carbon neutrality and the 1.5°C temperature increase threshold. Despite their knowledge of major greenhouse gases, scientists and policymakers struggle to accurately capture $CO_2$ emissions dynamics about their pathways (GHGs) (Magazzino et al., 2024). In this regard, we identified gaps in the literature regarding the analysis of the interactions of AI and nuclear energy technology budgeting (composite variable) with $CO_2$ emissions. Thus, this study employed a robust econometric approach in evaluating the present relationships between AI and nuclear energy technology budgeting (composite variable) with $CO_2$ emissions. On the methodological side, we introduce a novel approach by introducing moderating AI in the relationship between nuclear energy technology budgeting and $CO_2$ emissions. For empirical estimates, the study relies on the MM-QR method. Here, we were able to demonstrate that nuclear energy technology budgeting does influence all $CO_2$ emissions. We also observed that AI could contribute to $CO_2$ emissions directly. However, novel findings related to the interaction effect of AI and nuclear energy technology budget, the findings reveal that nuclear energy technology budgeting reduces carbon emissions through AI.

Policymakers may need to introduce incentives for renewable energy use in AI-related operations or promote the development of more energy-efficient AI algorithms. Governments could conduct thorough environmental impact assessments for AI initiatives, especially in industries that are heavy emitters. Efforts to develop AI solutions specifically targeted at reducing emissions and enhancing energy efficiency should be prioritized to counteract the possible environmental costs of AI deployment. Policymakers and stakeholders need to consider the short-term emissions increase associated with nuclear energy development against its long-term potential for low-emission energy production. Clear strategies can help bridge the gap until the emissions-reducing benefits of nuclear power are realized. Emissions during the construction and development phases could be minimized by adopting low-carbon materials, recycling construction waste, and utilizing renewable energy for construction activities. Investing in nuclear energy might be balanced with other renewable options, like solar and wind, that have shorter development times and can begin reducing emissions sooner. This integrated strategy could help mitigate the short-term emissions spike often linked to the development of nuclear energy. A positive coefficient may raise concerns among stakeholders or the public, who expect nuclear investments to lead to lower emissions. Therefore, it is essential to communicate effectively about nuclear project lifecycle emissions, including the initial increase in emissions and the long-term benefits. This communication is vital for managing expectations.

Policymakers and stakeholders should prioritize allocating funding and resources to integrate AI in nuclear energy projects, as this approach has the potential to enhance the environmental advantages of nuclear energy by reducing $CO_2$ emissions. Targeted research into AI applications for emission reduction can further improve its moderating effect. This could involve creating targeted AI algorithms to enhance energy efficiency, optimize carbon capture, and facilitate predictive maintenance designed explicitly for nuclear facilities. The partnership between the AI and nuclear energy sectors has the potential to foster innovations that leverage AI's capabilities to reduce emissions, establishing industry standards and frameworks for implementing AI in nuclear initiatives. Despite the long-term nature of nuclear projects, a negative moderating role of AI could accelerate the realization of emissions reduction benefits. Transparent reporting of these benefits may assist in managing public and stakeholder expectations, illustrating AI's role in enhancing the environmental viability of nuclear energy.

  The research highlights several significant limitations: obtaining reliable, real-time data on nuclear operations and budgets, as well as the adoption of AI in the nuclear sector, is often hindered by confidentiality and regulatory restrictions. This scarcity of data undermines the accuracy of AI model predictions and restricts the study's ability to reflect current technological and financial trends. Additionally, variations in data availability and accuracy across different regions pose substantial challenges in generalizing findings on a global scale. Furthermore, differing reporting standards among nations could affect the study's ability to consistently evaluate the impact of AI on nuclear energy investment and emissions reduction.

# References

Álvarez-herránz, A., Balsalobre, D., Cantos, J.M., Shahbaz, M., 2017. Energy Innovations-GHG Emissions Nexus : Fresh Empirical Evidence from crossmark. Energy Policy 101, 90–100. https://doi.org/10.1016/j.enpol.2016.11.030

Alvarez-herranz, A., Balsalobre-lorente, D., Shahbaz, M., 2017. Energy innovation and renewable energy consumption in the correction of air pollution levels. Energy Policy 105, 386–397. https://doi.org/10.1016/j.enpol.2017.03.009

Bashir, M.F., Shahbaz, M., Ma, B., Alam, K., 2024. Evaluating the roles of energy innovation, fossil fuel costs and environmental compliance towards energy transition in advanced industrial economies. J Environ Manage 351, 119709. https://doi.org/10.1016/j.jenvman.2023.119709

Beck, N., Katz, J.N., American, T., Science, P., 1995. Time-Series With Not To Do ) To Do ( and What Cross-Section. Am Polit Sci Rev 89, 634–647.

Chudik, A., Pesaran, M.H., 2022. An augmented Anderson–Hsiao estimator for dynamic short-T panels†. Econom Rev 41, 416–447. https://doi.org/10.1080/07474938.2021.1971388

Danish, Ulucak, R., 2022. Analyzing energy innovation-emissions nexus in China: A novel dynamic simulation method. Energy 244, 123010. https://doi.org/10.1016/j.energy.2021.123010

Danish, Ulucak, R., 2021. A revisit to the relationship between financial development and energy consumption: Is globalization paramount? Energy 227, 120337. https://doi.org/10.1016/j.energy.2021.120337

Fatouros, N., Stengos, T., 2023. Nuclear Energy, Economic Growth, and the Environment: Optimal policies in a model with endogenous technical change and environmental constraints. J Econ Asymmetries 28, e00325. https://doi.org/10.1016/j.jeca.2023.e00325

Hannah Ritchie and Max Roser, 2020. $CO_2$ emissions How much $CO_2$ does the world emit? Which countries emit the most? Published online at OurWorldinData.org. https://doi.org/Retrievedfrom'https://ourworldindata.org/co2-emissions'[OnlineResource]

Hassan, S.T., Batool, B., Wang, P., Zhu, B., Sadiq, M., 2023. Impact of economic complexity index, globalization, and nuclear energy consumption on ecological footprint: First insights in OECD context. Energy 263, 125628. https://doi.org/10.1016/j.energy.2022.125628

Hassan, S.T., Khan, D., Zhu, B., Batool, B., 2022. Is public service transportation increase environmental contamination in China? The role of nuclear energy consumption and technological change. Energy 238, 121890. https://doi.org/10.1016/j.energy.2021.121890

Huang, A., Dai, L., Ali, S., Sunday, T., 2024a. From funds to footprints : Unravelling the asymmetric association between nuclear energy technology and environmental quality. Energy 309, 133006. https://doi.org/10.1016/j.energy.2024.133006

Huang, A., Guo, M., Dai, L., Mirza, A., Ali, S., 2024b. Budgeting for a greener future: Asymmetric nexus between nuclear energy technology budgets and CO2 emissions. Technol Forecast Soc Change 202, 123321. https://doi.org/10.1016/j.techfore.2024.123321

IAEA, 2023. Enhancing Nuclear Power Production with Artificial Intelligence. Vienna. https://doi.org/https://www.iaea.org/bulletin/enhancing-nuclear-power-production-with-artificial-intelligence

IEA, 2022. World Energy Outlook 2022. https://doi.org/https://www.oecd.org/content/dam/oecd/en/publications/reports/2022/10/world-energy-outlook-2022_0989c209/3a469970-en.pdf

International Energy Agency, 2021. International Energy Agency (IEA), Renewables 2021: Analysis and forecast to 2026. International Energy Agency (IEA) Publications International. 2026.

Jessika Trancik, Patrick Brown, Joel Jean, Goksin Kavlak, Magdalena Klemun, Morgan Edwards, James McNerney, Marco Miotti, Joshua Mueller, Zachary Needell, 2015. MITEI-Technology-improvement-and-emissions-reductions-as-mutually-reinforcing-efforts.

Jin, X., Ahmed, Z., Pata, U.K., Kartal, M.T., Erdogan, S., 2023a. Do investments in green energy, energy efficiency, and nuclear energy R&D improve the load capacity factor? An augmented ARDL approach. Geoscience Frontiers 101646. https://doi.org/10.1016/j.gsf.2023.101646

Jin, X., Ahmed, Z., Pata, U.K., Kartal, M.T., Erdogan, S., 2023b. Do investments in green energy, energy efficiency, and nuclear energy R&D improve the load capacity factor? An augmented ARDL approach. Geoscience Frontiers 101646. https://doi.org/10.1016/j.gsf.2023.101646

Jordaan, S.M., Romo-Rabago, E., McLeary, R., Reidy, L., Nazari, J., Herremans, I.M., 2017. The role of energy technology innovation in reducing greenhouse gas emissions: A case study of Canada. Renewable and Sustainable Energy Reviews 78, 1397–1409. https://doi.org/10.1016/j.rser.2017.05.162

Magazzino, C., Cerulli, G., Haouas, I., Unuofin, J.O., Sarkodie, S.A., 2024. The drivers of GHG emissions: A novel approach to estimate emissions using nonparametric analysis. Gondwana Research 127, 4–21. https://doi.org/10.1016/j.gr.2023.10.004

Meckling, J., Galeazzi, C., Shears, E., Xu, T., Anadon, L.D., 2022. Energy innovation funding and institutions in major economies. Nat Energy 7, 876–885. https://doi.org/10.1038/s41560-022-01117-3

Nuclear Energy Agency, 2022. In this Issue…, NASSP Bulletin. https://doi.org/10.1177/01926365221125404

Obekpa, H.O., Alola, A.A., 2023. Asymmetric response of energy efficiency to research and development spending in renewables and nuclear energy usage in the United States. Progress in Nuclear Energy 156, 104522. https://doi.org/10.1016/j.pnucene.2022.104522

Pesaran, M.H., 2015. Testing Weak Cross-Sectional Dependence in Large Panels. Econom Rev 34, 1089–1117. https://doi.org/10.1080/07474938.2014.956623

Pesaran, M.H., 2007. A simple panel unit root test in the presence of cross sectional dependence. JOURNAL OF APPLIED ECONOMETRICS 22, 265–312. https://doi.org/DOI: 10.1002/jae.951

Renteria del Toro, F. de los A., Hao, C., Tokuhiro, A., Gomez-Fernandez, M., Gomez-Torres, A., 2024. Digitalization as an aggregate performance in the energy transition for nuclear industry. Nuclear Engineering and Technology 56, 1267–1276. https://doi.org/10.1016/j.net.2023.11.030

Sarkodie, S.A., Ahmed, M.Y., Owusu, P.A., 2023. Advancing COP26 climate goals: Leveraging energy innovation, governance readiness, and socio-economic factors for enhanced climate resilience and sustainability. J Clean Prod 431, 139757. https://doi.org/10.1016/j.jclepro.2023.139757

Shahbaz, M., Nasir, M.A., Roubaud, D., 2018. Environmental Degradation in France: The Effects of FDI, Financial Development, and Energy Innovations. Energy Econ #pagerange#. https://doi.org/10.1016/j.eneco.2018.07.020

Shen, S., Faridi, M.Z., Nazar, R., Ali, S., 2024. Asymmetric nexus between nuclear energy technology budgets and carbon emissions in European economies: Evidence from quantile-on-quantile estimation. Nuclear Engineering and Technology. https://doi.org/10.1016/j.net.2024.03.030

Sim, N., Zhou, H., 2015. Oil Prices , US Stock Return , and the Dependence Between Their Quantiles ∗. JOURNAL OF BANKING FINANCE. https://doi.org/10.1016/j.jbankfin.2015.01.013

Simionescu, M., 2023. The renewable and nuclear energy-economic growth nexus in the context of quality of governance. Progress in Nuclear Energy 157, 104590. https://doi.org/10.1016/j.pnucene.2023.104590

Stijn, B., 2023. Artificial intelligence and the labour market: Introduction.

Uche, E., Ngepah, N., Cifuentes-Faura, J., 2023. Upholding the green agenda of COP27 through publicly funded R&D on energy efficiencies, renewables, nuclear and power storage technologies. Technol Soc 75. https://doi.org/10.1016/j.techsoc.2023.102380

Wang, Q., Zhang, F., Li, R., Sun, J., 2024. Does artificial intelligence promote energy transition and curb carbon emissions? The role of trade openness. J Clean Prod 447, 141298. https://doi.org/10.1016/j.jclepro.2024.141298

Wang, Z., Zhu, Y., 2020. Do energy technology innovations contribute to CO2 emissions abatement? A spatial perspective. Science of The Total Environment 726, 138574. https://doi.org/10.1016/J.SCITOTENV.2020.138574

World Nuclear Association, 2014. Nuclear energy and sustainable development, World Nuclear Association. https://doi.org/10.1016/j.enpol.2014.10.001

Yıldırım, D.Ç., Esen, Ö., Yıldırım, S., 2022. The nonlinear effects of environmental innovation on energy sector-based carbon dioxide emissions in OECD countries. Technol Forecast Soc Change 182, 121800. https://doi.org/10.1016/J.TECHFORE.2022.121800

Yue, X., Peng, M.Y.P., Anser, M.K., Nassani, A.A., Haffar, M., Zaman, K., 2022. The role of carbon taxes, clean fuels, and renewable energy in promoting sustainable development: How green is nuclear energy? Renew Energy 193, 167–178. https://doi.org/10.1016/j.renene.2022.05.017

Zhan, L., Bo, Y., Lin, T., Fan, Z., 2021. Development and outlook of advanced nuclear energy technology. Energy Strategy Reviews 34, 100630. https://doi.org/10.1016/j.esr.2021.100630

Zhou, W., Zhang, Y., Li, X., 2024. Artificial intelligence, green technological progress, energy conservation, and carbon emission reduction in China: An examination based on dynamic spatial Durbin modeling. J Clean Prod 446, 141142. https://doi.org/10.1016/j.jclepro.2024.141142